\documentclass[preprintnumbers,floatfix,12pt,superscriptaddress,nofootinbib]{revtex4}

\usepackage{amsfonts}
\usepackage{amssymb,epsfig}
\usepackage{latexsym}
\usepackage{amsmath}
\usepackage{graphicx}

\begin{document}

\title{ Generalized holographic dark energy model described at the Hubble length}
\author{M. Malekjani \footnote{Email: \text{malekjani@basu.ac.ir}}}

 \address{Department of Physics, Faculty of Science, Bu-Ali Sina University, Hamedan 65178,
Iran}
\address{Research Institute for Astronomy and Astrophysics
of Maragha (RIAAM), Maragha, Iran\\}

\begin{abstract}
We generalize the holographic dark energy model described in Hubble
length IR cutoff by assuming a slowly time varying function for
holographic parameter $c^2$. We calculate the evolution of EoS
parameter and the deceleration parameter as well as the evolution of
dark energy density in this generalized model. We show that the
phantom line is crossed from quintessence regime to phantom regime
which is in agreement with observation. The evolution of
deceleration parameter indicates the transition from decelerated to
accelerated expansion. Eventually, we show that the GHDE with HIR
cutoff can interpret the pressureless dark matter era at the early
time and dark energy dominated phase later.

\end{abstract}
\maketitle

\newpage
Nowadays we are strongly believed that our universe experiences an
accelerated expansion. The complementary astronomical data gathered
from SNe Ia \cite{c1}, WMAP \cite{c2}, SDSS \cite{c3} and X-ray
\cite{c4} experiments confirm this cosmic acceleration. Within the
framework of general relativity (GR), a dark energy component with
negative pressure is introduced to explain this acceleration. Dark
energy scenario have got a lot of attention in modern cosmology. In
recent years a plenty theoretical models have been investigated to
interpret the dynamical properties of dark energy. One can see
\cite{rev1,rev2} for a review of dark energy models. The holographic
dark energy (HDE) model is one of these models to explain a dark
energy scenario. This model is constructed based on the holographic
principle in quantum gravity \cite{c11}. In quantum gravity, a short
distance ultra- violet (UV) cut-off is related to the long distance
infra-red (IR) cut-off, due to the limit set by the formation of a
black hole \cite{c11}. Based on the holographic principle, the
energy of a system with size $L$
 does not exceed the energy of black hole with the same
 size, i.e.,
\begin{equation}\label{black}
 L^3\rho_d\leq L m_{p}^2
 \end{equation}
where $m_{p}$ is the reduced plank mass. By saturating the
inequality (\ref{black}), the energy density of HDE model in
cosmology is identified by \cite{cohen11}
\begin{equation}\label{ircut}
 \rho_d=3c^2m_{p}^2L^{-2},
 \end{equation}
where $c^2$ is a numerical constant of order unity and the factor
$3$ was introduced for convenience. An interesting feature of HDE is
that it has a close connection with the space-time foam \cite{ng1}.
Another features of HDE model can be found in Section $3$ of
\cite{zimdeh}. From the observational point of view, the HDE model
has been constrained by various astronomical observation
\cite{obs3a,obs1,Wu:2007fs,obs3}. The HDE model has been also
investigated widely in the literature \cite{nonflat,holoext}. For a
recent review on different HDE models and their consistency check
with observational data see \cite{zimdeh2}. From the theoretical
point of view there are some motivations leading to the form of HDE
model \cite{mot2}. It should be noted that various HDE models have
been investigated via assuming different IR cutoffs. The simple
choice for IR cutoff is the Hubble radius, i.e., $L=H^{-1}$.  In
this case, the accelerated expansion of the universe can not be
achieved and we get a wrong equation of state for this model
\cite{c11}. However, in the presence of interaction between dark
matter and dark energy, the HDE model can derive the cosmic
acceleration and also, in this case, the cosmic coincidence problem
can be solved \cite{zimdeh5}. Event horizon is the another choice
for IR cutoff. Although, in this case the accelerated expansion can
be achieved, but the generalized second law (GSL) does not satisfy
in a universe enveloped by event horizon IR cutoff \cite{sheykhi33}.
The other choice for IR cutoff is the particle horizon. In this
case, the HDE model can not also obtain the late time accelerated
expansion \cite{c11}. Here same as \cite{sheykhi34} we assume Hubble
horizon as an IR cutoff for HDE model. In this case the GSL is also
satisfied in the interacting accelerating universe
\cite{sheykhi34}. Therefore the Hubble horizon is preferred from thermodynamical point of view.\\
 It is worthwhile to mention that the
parameter $c^2$ in HDE model has an essential role in characterizing
the properties of HDE model. For example, the HDE model can behave
as a phantom or quintessence dark energy models at the future for
the values of $c^2$ bigger or smaller than $1$,
respectively \cite{setar}.\\
In all above references the HDE model was assumed to have a constant
value for holographic parameter $c^2$. However there are no strong
evidences telling us that $c^2$ should be a constant parameter. In
general the term $c^2$ can be assumed as a function of time.By
slowly vary function with time, $\dot{(c^2)}/c^2$ is upper bounded
by the Hubble expansion rate, i.e.,
\begin{equation}
\frac{\dot{(c^2)}}{c^2}\leq H
\end{equation}
In this case the time scale of the evolution of $c^2$ is shorter
 than $H^{-1}$ and one can be satisfied to consider the time
 dependency of $c^2$ \cite{radi}. It has been also shown that the
 parameter $c^2$ can not be constant for all times during the
 evolution of the universe \cite{radi}.\\
As was mentioned above, in the presence of interaction between dark
matter and dark energy the HDE model with the Hubble horizon IR
cutoff can solve the the coincidence problem and late time
accelerated expansion. However, another alternative approach instead
of interaction between dark components is that the holographic
parameter $c^2$ varies slowly with time to solve the coincidence
problem and explain late time acceleration \cite{duran}.It has been
shown that the interacting model of dark
 energy in which the coincidence problem is alleviated can be recast
 as a noninteracting model in which the holographic parameter $c^2$
 evolves slowly with time \cite{duran}.\\
In the line of above studies, we consider the HDE model with
time-varying holographic parameter $c^2(z)$, namely: generalized
holographic dark energy (GHDE, hereafter). We also consider the
Hubble horizon as an IR cutoff (HIR, hereafter). We investigate the
EoS parameter of the model as well as the deceleration parameter and
discuss the density
evolution of dark energy in this model.\\

Let us start with flat Friedmann-Robertson-Walker (FRW) universe. In
this case the first Friedmann equation is given by
\begin{equation}\label{fridt}
H^{2}=\frac{1}{3m_{p}^{2}}(\rho _{m}+\rho _d)
\end{equation}%
where $\rho_m$ and $\rho_d$ are, respectively, the energy densities
of pressureless dark matter and dark energy and $m_p$ is the reduced
planck mass. For Hubble radius IR cutoff, $L=H^{-1}$, the energy
density of GHDE model from (\ref{ircut}) can be given by
\begin{equation}\label{dens1}
\rho_d=3m_p^2c^2(z)H^2
\end{equation}
where the holographic parameter is considered as a function of
redshift.\\
We now define the dimensionless energy density parameters as
\begin{equation}\label{denergyt}
\Omega_{m}=\frac{\rho_m}{\rho_c}=\frac{\rho_m}{3M_p^2H^2}, ~~~\\
\Omega_d=\frac{\rho_d}{\rho_c}=\frac{\rho_d}{3M_p^2H^2}=c^2(z)~~\\
\end{equation}
According to these definitions, the first Friedmann equation in
spatially flat universe can be written as follows
\begin{equation}
\Omega _{m}+\Omega _{\Lambda}=1.  \label{Freqt}
\end{equation}%
The conservation equations for pressureless dark matter and dark
energy, respectively, are given by
\begin{eqnarray}
\dot{\rho _{m}}+3H\rho _{m}=0, \label{contmt}\\
\dot{\rho _d}+3H(1+w_d)\rho_d=0. \label{contdt}
\end{eqnarray}%
Taking the time derivative of Friedmann equation (\ref{fridt}) and
using (\ref{Freqt}, \ref{contmt}, \ref{contdt}), one can obtain
\begin{equation}\label{hdott}
\frac{\dot{H}}{H^2}=-\frac{3}{2}[1+w_{\Lambda}\Omega_d]
\end{equation}
Also it is obvious to see that differentiating Eq.(\ref{dens1}) with
respect to time yields
\begin{equation}\label{dotdens}
\dot{\rho_d}=2\rho_d(\frac{\dot{c}}{c}+\frac{\dot{H}}{H})
\end{equation}
Inserting (\ref{dotdens}) and (\ref{dens1}) in conservation equation
for dark energy (\ref{contdt}) and using (\ref{hdott}), we find the
equation of state, $w_d$, for GHDE model with Hubble length as
\begin{equation}\label{eos1}
w_d=-\frac{2c^{\prime}}{3c(1-c^2)}
\end{equation}
where prime represents the derivative with respect to $\ln{a}$. It
is clear that the above relation reduce to $w_d=0$ for constant
holographic parameter $c$. Hence, as expected, the HDE model in HIR
gets to wrong equation of state for dark energy which can no
describe the expanding universe. As was mentioned before, this
problem for HDE model can be solved, if we consider the interaction
between dark matter and dark energy (see \cite{sheykhi34} for more
detail). In is worthwhile to mention that from (\ref{eos1}) one can
see that the GHDE model in which the holographic parameter $c$ is
considered as a function of redshift can get $w_d<0$ in HIR without
assuming the interaction parameter. For this aim we use the
Wetterich parametrization in which the holographic parameter $c$ is
considered in terms of redshift as follows \cite{witt}
\begin{equation}\label{witt}
c(z)=\frac{c_0}{1+c_1\ln{(1+z)}}
\end{equation}
Putting $c_1=0$, the above holographic parameter reduces to $c=c_0$
indicating the constant value for HDE model. At the present time:
$z\rightarrow 0$, $c(z)\rightarrow c_0$, and at the early
time:$z\rightarrow \infty$, $c(z)\rightarrow 0$. Hence the
holographic parameter varies slowly from zero to $c_0$ during the
history of the universe. Also to have positive energy density for
dark energy, $\rho_d\geq 0$, we should take the condition: $c_0>0$
and $c_1\geq0$. In numerical procedure, we chose matter density
parameter $\Omega_m=1-c_0^2$, and dark energy density parameter
$\Omega_d=c_0^2$, indicating the spatially flat universe. In
Fig.(1), by solving (\ref{eos1}) and using (\ref{witt}), we plot the
evolution of EoS parameter, $w_d$, in terms of redshift $z$ for
different illustrative values of $c_0$ and $c_1$. Here we see that
the EoS parameter, $w_d$ of GHDE with HIR can transit from
quintessence regime ($w_d>-1$) to phantom regime ($w_d<-1$). The
observations favor dark energy models which cross the phantom line
$w=-1$ from up ($w_d<-1$) to down ($w_d<-1$) in near past
\cite{obs3}. Therefore this model is compatible with observations.
Contrary the GHDE model HIR, the EoS parameter for interacting HDE
model with HIR is constant during the history of the universe ( see
Eq.(8) of \cite{sheykhi34}). However, Sheykhi showed that by
applying some restrictions on the interaction parameter $b$ and
model parameter $c$, the EoS parameter of the HDE model with HIR can
behave as a quintessence or a phantom type dark energy. But, neither
the quintessence nor the phantom alone can fulfill the transition
from $w_d>-1$ to $w_d<-1$.\\
We now calculate the evolution of energy density of GHDE model
described at HIR. From Eq. (\ref{denergyt}), the dark energy density
of GHDE equals to square of varying holographic parameter,
$\Omega_d=c^(z)$. Using (\ref{witt}), in Fig.(2) the evolution of
dark energy density is plotted in terms of redshift for some
illustrative values of model parameters $c_0$ and $c_1$. At the
early time ($z \rightarrow\infty)$ the parameter
$\Omega_d\rightarrow 0$ which represents the dark matter dominated
universe at the early time. Then the parameter $\Omega_d$ increases
to its present value $c_0$ which indicates the dark energy dominated
epoch. The important note is that in standard HDE model under HIR
cutoff, since the model parameter $c$ is constant therefore the
energy density $\Omega_d=c^2$ has no evolution during the history of
the universe, i.e., $\Omega_d$ is constant from early time to
present time. Unlike the standard HDE model, in GHDE model with HIR
the parameter $\Omega_d$ increases from zero at the early time and
tends to its present value at the present epoch. This behavior of
GHDE model can interpret the decelerated expansion at the early time
dark matter dominated universe and also the accelerated expansion at
the dark energy dominated epoch.\\
For completeness, we calculate the deceleration parameter $q$ for
GHDE model with HIR. The positive value of deceleration parameter
($q>0$) indicates the decelerated phase of expansion and the
negative value ($q<0$) represents the accelerated phase of expansion
of the universe. The parameter $q$ is defined as
\begin{equation}\label{q1}
q=-1-\frac{\dot{H}}{H^2}
\end{equation}
Inserting (\ref{hdott}) in (\ref{q1}) and using (\ref{eos1}) as well
as $\Omega_d=c^2$, the parameter $q$ for GHDE with HIR can be
obtained as
\begin{equation}\label{q2}
q=\frac{1}{2}-\frac{c^{\prime}c}{1-c^2}
\end{equation}
In the limiting case of standard HDE model  with constant value of
$c$, the parameter $q$ reduces to $q=1/2$ which describes the
decelerated phase and can not represents the accelerated phase.
However, including the interaction parameter and  some restrictions
on the interaction parameter $b$ and model parameter $c$ in standard
HDE model with HIR can result the negative value for deceleration
parameter $q$ \cite{sheykhi34}, but since the parameter $c$ is
constant therefore the transition from decelerated to accelerated
expansion can not be achieved.\\
In Fig.(3), by solving (\ref{q2}) and using (\ref{witt}), we plot
the parameter $q$ in GHDE with HIR model as a function of redshift
parameter $z$. We see that at the early times the parameter $q$ is
$1/2$ indicating the decelerated phase at dark matter-dominated
universe. Then the parameter $q$ reaches to negative values
representing the accelerated phase at dark energy-dominated
background. This property of GHDE with HIR is consistent with this
observational fact that the universe has entered to the accelerated
phase at past times \cite{obs6}.\\

In summery, we considered the generalized holographic dark energy
model in spatially flat universe described in Hubble length an an IR
cutoff (GHDE with HIR cutoff). The holographic parameter $c$
generally is not constant and can be assumed as a function of cosmic
redshift. The standard HDE model described by HIR cutoff gets to
wrong equation of state for dark nergy \cite{c11}. The observations
favor dark energy models which cross the phantom line $w=-1$ from
quintessence regime ($w_d<-1$) to phantom regime ($w_d<-1$) in near
past \cite{obs3} and also the models in which the deceleration
parameter transit from positive value to negative value \cite{obs6}.
Although, including the interaction between dark matter and dark
energy in standard HDE model described by HIR cutoff can solve the
coincidence problem and late time accelerated expansion
\cite{sheykhi33,sheykhi34}, but the EoS parameter of this model
behaves as a quintessence or phantom model and can not transit from
quintessence regime ($w_d>-1$) to phantom regime
($w_d<-1$)\cite{sheykhi34}. Also in the context of interacting HDE
with HIR cutoff the deceleration parameter $q$ is negative for all
times in the history of the universe and therefore can not explain
the transition from decelerated to accelerated expansion. However,
in the case of GHDE with HIR cutoff, we obtained the EoS parameter
as well as the deceleration parameter and evolution of dark energy
density. Here we assumed the holographic parameter $c^2$ varies
slowly with time instead of adding the interaction term. We showed
that in this model the phantom line is crossed from up ($w_d>-1$) to
down ($w_d<-1$) which is in agreement with observation \cite{obs3}.
Also it has been shown that in this model the evolution of
deceleration parameter $q$ indicates the decelerated phase at the
early time ($q>0$) and accelerated phase at later ($q<0$). The
evolution of energy density of dark energy represents the
pressureless dark matter-dominated universe at the early time and
dark energy-dominated phase at the present time.

\begin{figure}[!htb]
\includegraphics[width=8cm]{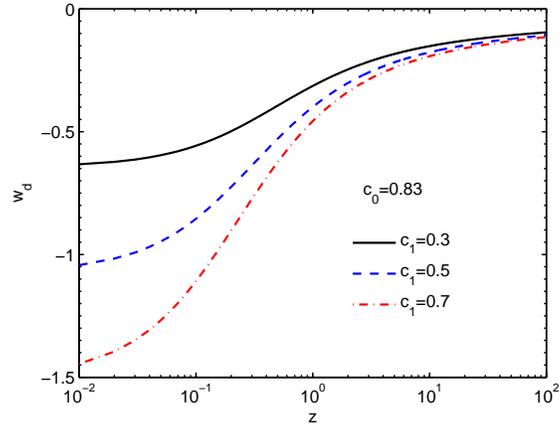}
\caption{The evolution of EoS parameter of GHDE model with HIR
cutoff versus redshift parameter $z$ for different values of model
parameters $c_0$ and $c_1$. Here we take $\Omega_m^0=1-c_0^2$ and
$\Omega_d^0=c_0^2$. }
\end{figure}

\begin{figure}[!htb]
\includegraphics[width=8cm]{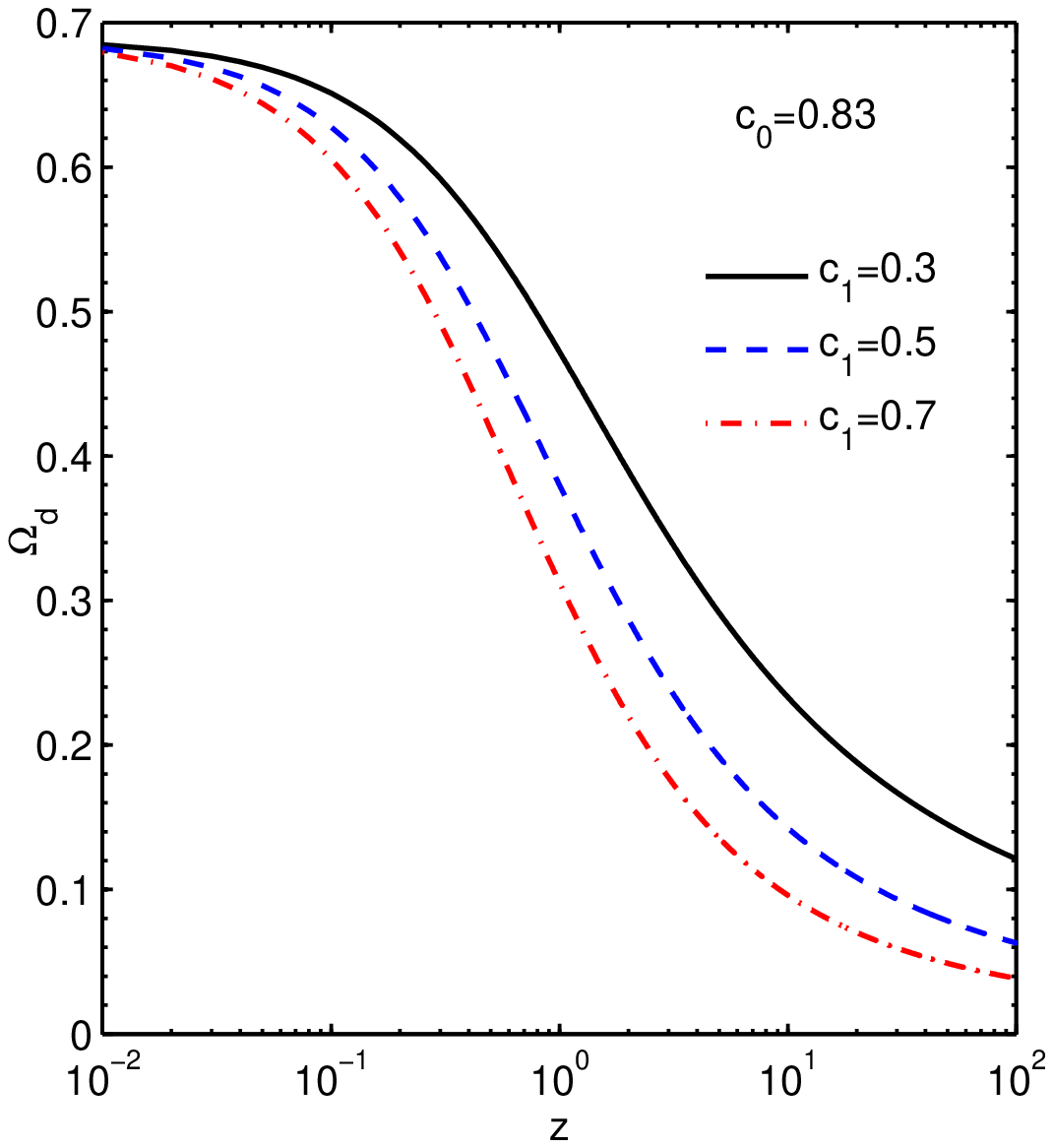} \caption{The evolution of energy density of GHDE model with
HIR cutoff versus redshift parameter $z$ for different values of
model parameters $c_0$ and $c_1$. Here we take $\Omega_m^0=1-c_0^2$
and $\Omega_d^0=c_0^2$. }
\end{figure}

\begin{figure}[!htb]
\includegraphics[width=8cm]{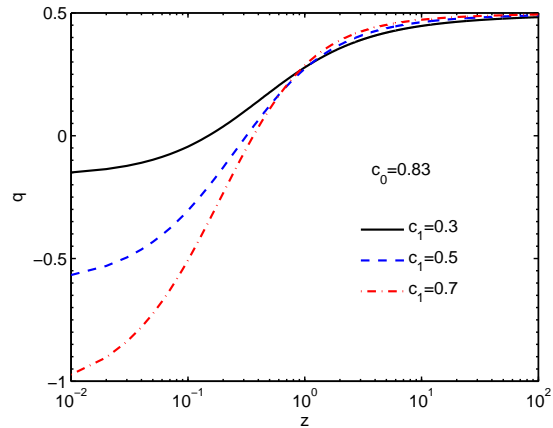} \caption{The evolution of deceleration parameter $q$ in GHDE with
HIR model versus redshift parameter $z$ for different illustrative
values of model parameters $c_0$ and $c_1$. Here we take
$\Omega_m^0=1-c_0^2$ and $\Omega_d^0=c_0^2$. }
\end{figure}

\newpage\

\end{document}